# Imry-Ma phase in *O(n)* models for space dimensions higher than the lower critical dimensionality

A.A. Berzin[a], A.I. Morosov[b]*, and A.S. Sigov[a]

[a]MIREA - Russian Technological University, 78 Vernadskiy Ave., 119454 Moscow, Russian Federation

[b]Moscow Institute of Physics and Technology (National Research University), 9 Institutskiy per., 141700 Dolgoprudny, Moscow Region, Russian Federation

* e-mail: mor-alexandr@yandex.ru

Systems with continuous symmetry of the vector order parameter containing defects of the "random local field" or "random local anisotropy" types are investigated. It is shown that the disordered Imry-Ma phase, in which the order parameter follows the spatial fluctuations of the random field or random anisotropy direction, can also occur in a coordinate space of dimension $d$ higher than the lower critical dimension $d_l = 4$. For this, the hyperplanes of dimension $m < d_l$, must exist in the coordinate space in which there is a strong exchange interaction between spins, and the interaction between spins belonging to adjacent hyperplanes should not exceed a certain critical value.

**Key words:** defects of the "random local field" and "random local anisotropy" types; *O(n)* models; critical dimensionality; Imry-Ma phase.

## 1. Introduction

In their pioneering work [1], Imry-Ma, examining systems with continuous symmetry of the *n*-component vector order parameter (*O (n)* models) containing defects of the "random local field" type, found the lower critical dimension of the coordinate space $d_l = 4$. In spaces of a dimension $d < d_l$, the long-range order is absent in the ground state of the system, and the disordered Imry-Ma phase is realized in which the order parameter follows large-scale static fluctuations of the field created by the defects. In spaces of a dimension $d > d_l$, the ground state is the state with the long-range order.

This conclusion is made within the framework of a model in which the magnitude of the exchange interaction in all directions in the *d*-dimensional space is the same. Below, we restrict ourselves to the approximation of the nearest neighbor interaction and consider a rectangular *d*-dimensional lattice in which the distance between the nearest neighbors $a_1$ in *m* directions is less than the similar distance $a_2$ in the remaining *d-m* directions. This leads to the fact that the exchange integral $J_\parallel \equiv J(a_1)$ far exceeds the exchange integral $J_\perp \equiv J(a_2)$. It will be shown that if the value of $J_\perp$ turns out to be less than a certain critical value, then the Imry-Ma phase can also arise in a space of dimension $d > d_l$.

## 2. System of classical spins

The exchange interaction energy of *n*-component localized spins $\mathbf{s}_{i,\alpha}$ of a fixed unit length (the length of the vector can be included in the corresponding interaction constants or fields) in the framework of the model described above has the form

$$W_{ex} = -\frac{1}{2}J_\parallel \sum_{i,\alpha,k} \mathbf{s}_{i,\alpha}\mathbf{s}_{i+k,\alpha} - \frac{1}{2}J_\perp \sum_{i,\alpha,\delta} \mathbf{s}_{i,\alpha}\mathbf{s}_{i,\alpha+\delta}, \qquad (1)$$

where $J_\parallel, J_\perp > 0$, the index *i* numbers the spins in a hyperplane of dimension *m* with a strong exchange interaction of spins, the index *α* is the number of the hyperplane, the summation over *k* is carried out over the nearest neighbors to the given spin in the same hyperplane, and the summation over $\delta$ is performed over the neighbors closest to the given spin, lying in the neighboring hyperplanes.

The energy of interaction of spins with random local fields of defects is

$$W_{def} = -\sum_l \mathbf{s}_{i_l,\alpha_l} \mathbf{h}_{i_l,\alpha_l}, \qquad (2)$$

the summation is carried out over defects randomly located at the lattice sites, $\mathbf{h}_{i_l,\alpha_l}$ is the local field of the $l$-th defect, and the distribution density of random local fields $\mathbf{h}$ in the spin space (the space of the order parameter) is isotropic, which ensures the absence of the mean field in an infinite system and the effective anisotropy induced by the fields of defects [2].

### 3. Imry-Ma arguments

It is natural to assume that the characteristic size of the inhomogeneity of the order parameter $L_\parallel$ measured in the units of the corresponding lattice constant in the hyperplane with a strong exchange interaction is much larger than the similar size $L_\perp$ in the directions with a weak exchange interaction. The characteristic volume of the region in which we consider the statistical fluctuation of the random field of the defects is $L_\parallel^m L_\perp^{d-m}$. For the given region, the value of the average field $\widetilde{\mathbf{h}}$, which arose due to the predominance of defects with one direction of the field, is in the order of magnitude equal to

$$\widetilde{\mathbf{h}} = \left(\frac{c\langle \mathbf{h}_{i_l,\alpha_l}^2 \rangle}{L_\parallel^m L_\perp^{d-m}}\right)^{1/2}, \qquad (3)$$

here the angular brackets indicate averaging over the fields of all defects. The energy gain in comparison with the homogeneous ferromagnetic ground state (per a unit cell) due to the order parameter following the average field in each such region is $w_{def} \approx -\widetilde{\mathbf{h}}$.

The energy of heterogeneous exchange per a unit cell is equal to

$$w_{ex} \approx \frac{mJ_\parallel}{L_\parallel^2} + \frac{(d-m)J_\perp}{L_\perp^2}. \qquad (4)$$

By minimizing the total energy $w = w_{def} + w_{ex}$ with respect to $L_\parallel$, we obtain the optimum values $L_\parallel^*$ и $w^*$:

$$L_\parallel^* \approx \left(\frac{16 J_\parallel^2 L_\perp^{d-m}}{c \langle \mathbf{h}_{i_l, \alpha_l}^2 \rangle}\right)^{\frac{1}{4-m}}, \qquad (5)$$

$$w^* \approx -(4-m)\left(\frac{c \langle \mathbf{h}_{i_l, \alpha_l}^2 \rangle}{16 J_\parallel^{m/2} L_\perp^{d-m}}\right)^{\frac{2}{4-m}} + \frac{(d-m) J_\perp}{L_\perp^2}. \qquad (6)$$

We restrict ourselves to the case of weak random fields $\langle \mathbf{h}_{i_l, \alpha_l}^2 \rangle \ll J_\parallel^2$. For the existence of the Imry-Ma phase, it is necessary that the dimension of a hyperplane with a strong exchange interaction $m$ be less than the lower critical dimension $d_l$=4. In this case, we have $L_\parallel^* \gg 1$.

In order for expression (6) to have a local minimum with respect to the variable $L_\perp$, it is necessary and sufficient that $\frac{2(d-m)}{4-m} < 2$. This gives the well-known Imry and Ma condition: $d < 4$. Otherwise, for $d \geq 4$, the value $w^*$ reaches the minimum value either for $L_\perp = 1$ or for $L_\perp \to \infty$. The infinite value of $L_\perp$ corresponds to the long-range ferromagnetic order, and $w^*(L_\perp \to \infty) = 0$. The value $L_\perp = 1$ corresponds to the Imri-Ma state with no correlation between spins belonging to different hyperplanes introduced above. In order for the Imry-Ma state to be the ground one, it should be met the inequality

$$w^*(L_\perp = 1) \approx -(4-m)\left(\frac{c \langle \mathbf{h}_{i_l, \alpha_l}^2 \rangle}{16 J_\parallel^{m/2}}\right)^{\frac{2}{4-m}} + (d-m) J_\perp < 0, \qquad (7)$$

which gives a restriction on the value of the smaller of the exchange integrals

$$\eta \equiv \frac{J_\perp}{J_\parallel} < \frac{4-m}{d-m}\left(\frac{c \langle \mathbf{h}_{i_l, \alpha_l}^2 \rangle}{16 J_\parallel^2}\right)^{\frac{2}{4-m}}. \qquad (8)$$

For $d$=4, $c \sim 10^{-1}$, and $\frac{\langle \mathbf{h}_{i_l, \alpha_l}^2 \rangle}{J_\parallel^2} \sim 10^{-2}$ we obtain $\eta < 10^{-3}$, $\eta < 10^{-4}$, and $\eta < 10^{-8}$ for $m = 1, 2, 3$ respectively.

## 4. Individual pinning of the order parameter by the fields of defects

For the applicability of the above estimates, it is necessary that for $L_\perp = 1$ the inequality

$$c(L_\parallel^*)^m \gg 1 \qquad (9)$$

should be valid, that is, the fluctuation region with optimal sizes should contain a large number of defects. For weak random fields, this condition is automatically satisfied for $m = 2; 3$. For $m = 1$, the inequality (9) leads to the condition

$$cJ_\parallel \gg \sqrt{\langle \mathbf{h}_{i_l,\alpha_l}^2 \rangle}. \qquad (10)$$

In the case of the reverse inequality

$$cJ_\parallel \ll \sqrt{\langle \mathbf{h}_{i_l,\alpha_l}^2 \rangle}, \qquad (11)$$

the effect of defects of the "random local field" type on the order parameter ceases to be collective. For $cL_\parallel^* < 1$, a situation is realized in which, in a one-dimensional spin chain with a strong exchange interaction, the order parameter unfolds from defect to defect so that at each defect it is directed along its random field. In this case, the characteristic value $L_\parallel$ is of the order of $c^{-1}$, the energy of the order parameter interaction with defects is

$$w_{def} \approx -c\sqrt{\langle \mathbf{h}_{i_l,\alpha_l}^2 \rangle}, \qquad (12)$$

and the total energy of the Imry-Ma phase per a unit cell is

$$w^* \approx -c\sqrt{\langle \mathbf{h}_{i_l,\alpha_l}^2 \rangle} + c^2 J_\parallel + (d-1)J_\perp. \qquad (13)$$

Spins belonging to different chains remain uncorrelated.

Since the second term on the right-hand side of expression (13) is much smaller in magnitude than the first one, the condition for the existence of the Imry-Ma phase in the case of the validity of inequality (11) has the form

$$J_\perp < \frac{c\sqrt{\langle \mathbf{h}^2_{i_l, \alpha_l}\rangle}}{d-1}. \tag{14}$$

For $d = 4$, $c \sim 10^{-2}$, and $\frac{\langle \mathbf{h}^2_{i_l, \alpha_l}\rangle}{J_\parallel^2} \sim 10^{-1}$, one finds $\eta < 10^{-3}$.

Otherwise, the ground state of the system is the phase with the long-range ferromagnetic order.

### 5. Transition temperature to the Imry-Ma phase

It was stated in Ref. [3] that the temperature $\tilde{T}$ of the transition from the paramagnetic phase to the Imry-Ma phase can be estimated from the condition $L_\parallel^* \sim r_c$, where $r_c$ is the correlation radius of the order parameter of the pure subsystem, in our case, a hyperplane with strong interaction of spins. At $T > \tilde{T}$, the condition $r_c < L_\parallel^*$ is fulfilled, and dynamic thermal fluctuations of the order parameter characteristic of a pure system are observed in the system. For $T < \tilde{T}$, the inequality $r_c > L_\parallel^*$ is valid and the static fluctuations of random field "freeze" these dynamic fluctuations. There exists a phenomenon akin to the glass transition phenomenon.

In the case $m = 3$, in a defect-free hyperplane, a second-order phase transition to the ferromagnetic phase occurs, when approaching this transition, the correlation radius of the order parameter tends to infinity [4]. Based on this, we can conclude that $\tilde{T}$ is of the order of the phase transition temperature of the defect-free three-dimensional *O(n)* model.

In the case of the *X-Y* model ($n = 2$) and a two-dimensional defect-free hyperplane ($m = 2$), a transition from the paramagnetic phase with an exponential character of a decrease in the correlation function to the Berezinsky-Kosterlitz-Thoules (BKT) phase with a power-law character of the decrease in the correlation function occurs in it [5, 6]. Since the correlation radius diverges when approaching the transition point, $\tilde{T}$ is of the order of the BKT transition temperature for the two-dimensional *O(2)* model $T_{BKT} = \pi J_\parallel/2$ [7].

In the remaining cases, when $m = 1$ or $m = 2$ and $n \geq 3$, the long-range order in the defect-free system occurs only at absolute zero temperature. The correlation radius diverges as the temperature approaches this value (see, for example, data on the two-dimensional Heisenberg model [7], the one-dimensional *X-Y* model [8, 9], and the one-dimensional Heisenberg model [10, 11]). Summarizing the calculation made in Ref. [10], we obtained the expression for the correlation radius of the one-dimensional system with the n-component order parameter ($m=1$; $n \geq 2$)

$$r_c = \frac{\gamma J_\parallel}{T} \equiv \gamma a, \tag{15}$$

where

$$\gamma = \lim_{a \to \infty} \left[ a \ln \frac{\int_0^\pi \cos\theta \exp(a \cos\theta) \sin^{n-2}\theta \, d\theta}{\int_0^\pi \exp(a \cos\theta) \sin^{n-2}\theta \, d\theta} \right]^{-1} = \frac{2}{n-1}. \tag{16}$$

Since the value $L_\parallel^*$ is finite, the transition to the Imry-Ma phase occurs at a finite nonzero temperature. The corresponding $\tilde{T}$ values are given in the Table.

## 6. Defects of "random local anisotropy" type

In the case of *O(n)* models containing point defects of the "random local anisotropy" type, the energy of interaction of defects with the order parameter is described by the expression

$$W_{imp} = -K_0 \sum_l (\mathbf{s}_l \mathbf{n}_l)^2, \tag{17}$$

where $K_0 > 0$ is the defect induced anisotropy constant, the summation is performed over defects randomly located at the lattice sites, and $\mathbf{n}_l$ is a unit vector that sets the direction of a random easy axis. The procedure for finding the conditions for the appearance of the Imry-Ma phase in such models is completely analogous to that carried out for systems with defects of the "random local field" type. The change reduces to replacing $\langle \mathbf{h}_{i_l, \alpha_l}^2 \rangle$ with $K_0^2$ in the final formulae.

## 7. Conclusion

The main conclusions of the work are formulated below.

1. In systems with continuous symmetry of the *n*-component vector order parameter containing defects of the "random local field" or "random local anisotropy" types, the disordered Imry-Ma phase may appear even if the space dimension exceeds the lower critical dimension $d_l = 4$ obtained in the case of identical exchange interaction with all nearest neighbors.

2. For the appearance of the Imry-Ma phase, it is necessary that in the coordinate space there exist hyperplanes of a dimension lower than $d_l$ with a strong exchange interaction of spins, and the exchange interaction between neighboring spins belonging to different hyperplanes should not exceed a critical value.

# References

1. Y. Imry and S.-k. Ma. Phys. Rev. Lett. **35**, 1399 (1975).
2. A. A. Berzin, A. I. Morosov, A. S. Sigov, Phys. Solid State **59**, 2016 (2017).
3. A. A. Berzin, A. I. Morosov, A. S. Sigov, Phys. Solid State 62, 332 (2020).
4. L. D. Landau and E. M. Lifshitz. *Course of Theoretical Physics*, Vol. 5: *Statistical Physics* (Butterworth–Heinemann, Oxford, 1980).
5. V. L. Berezinskii, Sov. Phys. JETP **32**, 493 (1971); **34**, 610 (1972).
6. J. M. Kosterlitz and D. G. Thouless. J. Phys. C **6**, 1181 (1973).
7. Yu. A. Izyumov, Yu. N. Skryabin. Statistical mechanics of magnetically ordered systems. Nauka, Moscow (1987). 268 c. (in Russian).
8. F. Wegner. Zeitschrift fur Physik **206**, 465 (1967).
9. R. W. Gerling and D. P. Landau. Phys. Rev. B **27**, 1719 (1983).
10. M. E. Fisher. Amer. J. Phys. **32**, 343 (1964).
11. D. J. Scalapino, Y. Ymry and P. Pincus. Phys. Rev. B **11**, 2042 (1978).

Table. Evaluation of the temperature for the Imry-Ma phase initiation.

| Model | $L_\parallel^*, (L_\parallel)$ | $r_c$ | $\tilde{T}$ |
|---|---|---|---|
| $m=2; n=3$ | $\left(\dfrac{16 J_\parallel^2}{c \langle \mathbf{h}_{i_l,\alpha_l}^2 \rangle}\right)^{1/2}$ | $\exp\left(\dfrac{2\pi J_\parallel}{T}\right)$ | $\dfrac{4\pi J_\parallel}{\ln\left(\dfrac{J_\parallel^2}{c \langle \mathbf{h}_{i_l,\alpha_l}^2 \rangle}\right)}$ |
| $m=1; n\geq 2;$ $cJ_\parallel \gg \sqrt{\langle \mathbf{h}_{i_l,\alpha_l}^2 \rangle}$ | $\left(\dfrac{16 J_\parallel^2}{c \langle \mathbf{h}_{i_l,\alpha_l}^2 \rangle}\right)^{1/3}$ | $\dfrac{2 J_\parallel}{(n-1)T}$ | $\dfrac{1}{n-1}\left(\dfrac{cJ_\parallel \langle \mathbf{h}_{i_l,\alpha_l}^2 \rangle}{2}\right)^{1/3}$ |
| $m=1; n\geq 2;$ $cJ_\parallel \ll \sqrt{\langle \mathbf{h}_{i_l,\alpha_l}^2 \rangle}$ | $c^{-1}$ | $\dfrac{2 J_\parallel}{(n-1)T}$ | $\dfrac{2cJ_\parallel}{n-1}$ |